\documentclass[prb,amsmath,twocolumn,superscriptaddress,showpacs]{revtex4}

\usepackage{amsfonts}
\usepackage{amsmath}
\usepackage{amssymb}
\usepackage{graphicx}
\usepackage{color}
\usepackage{bm}
\usepackage{enumerate}

\newcommand{\nn}{\nonumber}

\def\be{\begin{equation}}
\def\ee{\end{equation}}
\def\bea{\begin{eqnarray}}
\def\eea{\end{eqnarray}}
\def\bsp{\begin{split}}
\def\esp{\end{split}}

\begin{document}
\title{Re-summation of fluctuations near ferromagnetic quantum critical points}
\author{C.~J. Pedder}
\email{c.pedder@ucl.ac.uk}
\affiliation{London Centre for Nanotechnology, University College London, 17-19 Gordon St, London, WC1H 0AH, UK}
\author{F. Kr\"uger}
\email{frank.kruger@st-andrews.ac.uk}
\affiliation{SUPA, Department of Physics, University of St. Andrews, North Haugh, St Andrews, Fife. KY16 9SS}
\author{A.~G. Green}
\email{andrew.green@ucl.ac.uk}
\affiliation{London Centre for Nanotechnology, University College London, 17-19 Gordon St, London, WC1H 0AH, UK}

\date{\today}

\begin{abstract} 
We present a detailed analysis of the non-analytic structure of the free energy for the itinerant ferromagnet near the quantum critical point in two and three dimensions. We analyze a model of electrons with an isotropic dispersion interacting through a contact repulsion. A fermionic version of the quantum order-by-disorder mechanism allows us to calculate the free energy as a functional of the dispersion in the presence of homogeneous and spiralling magnetic order. We re-sum the leading divergent contributions, to derive an algebraic expression for the non-analytic contribution to free energy from quantum fluctuations. Using a recursion which relates sub-leading divergences to the leading term, we calculate the full $T=0$ contribution in $d=3$. We propose an interpolating functional form, which allows us to track phase transition lines at temperatures far below the tricritical point and down to $T=0$. In $d=2$, quantum fluctuations are stronger and non-analyticities more severe. Using a similar re-summation approach, we find that despite the different non-analytic structures, the phase diagrams in two and three dimensions are remarkably similar, exhibiting an incommensurate spiral phase near to the avoided quantum critical point. 
\end{abstract}


\maketitle

\section{Introduction}

Quantum fluctuations can have a dramatic effect on the phase behavior of the itinerant ferromagnet in the vicinity of the $T=0$ quantum critical point. 
In a pioneering paper,\cite{hertz1976quantum} Hertz suggested that fluctuations can lead to new scaling in the physical quantities above the 
quantum critical point, which he termed ``quantum critical scaling''.  The theory of ferromagnetic quantum criticality was further developed by 
Moriya\cite{Moriya85} and Millis.\cite{millis} More recently, several authors \cite{belitz1,Belitz+99,andre2, joe1, catherine1,joe2, andre1} noted that the coupling 
of soft electronic particle-hole modes to the magnetic order parameter generates non-analytic terms in the free energy that render the magnetic 
transitions first-order at low temperatures. Such fluctuation-induced first-order behavior is expected for ferromagnets in the Ising, XY, and 
Heisenberg universality classes in dimensions $1<d\le 3$, regardless of whether the moments are supplied by the conduction electrons or by electrons 
in another band.\cite{belitzkirkpatrick} Such first-order behaviour is also stable against weak disorder.\cite{belitz1,bkv} This explains why discontinuous ferromagnetic transitions 
are seen in various materials, including MnSi (Ref.~\onlinecite{mnsiexp}), $\textrm{Sr}_{1-x} \textrm{Ca}_x \textrm{Ru} \textrm{O}_3$ (Ref.~\onlinecite{ruthexp}), 
CoO$_2$ (Ref.~\onlinecite{Otero+08}), $\textrm{U} \textrm{Ge}_2$ (Ref.~\onlinecite{uge2exp}), and URhGe (Ref.~\onlinecite{urugeexp}).

It has been argued\cite{andre2} that fluctuation-driven first-order transitions are indicative of the appearance of precursor, incommensurate magnetic states. 
Recently, this possibility has been explored in $d=3$ within a fermionic version of the quantum order-by-disorder mechanism.\cite{andrewgareth,una,Kruger+12}
The central idea is to self-consistently calculate fluctuations around magnetically ordered states. This approach not only reproduces the non-analytic 
free energy of the homogeneous ferromagnet within a much simpler calculation, but also predicts the formation of an incommensurate spiral state, pre-empting 
the first-order transition into the ferromagnet. Spiral formation may be driven in mean-field by e.g. a feature in the density of states \cite{BGGS,BGSG,IZKI,KYI,IIK} or a breaking of inversion symmetry 
by the crystal structure, resulting in a Dzyaloshinski-Moriya interaction. \cite{Moriya, Dzyaloshinsky, BJ} Our mechanism for spiral formation does not require either of these complications, and is driven by quantum fluctuations of intinerant electrons with a simple, isotropic dispersion.

The quantum order-by-disorder phenomenon has much in common with the Coleman-Weinberg mechanism of mass generation in high energy 
physics\cite{coleman} and with the Casimir effect.\cite{casimir} It is familiar in the context of condensed matter physics.\cite{villain,Chandra+90,Mila+91,
Zaanen00,Kruger+06} However, where the fermionic version differs from all these approaches is that it encodes the quantum fluctuations in 
fermionic particle-hole excitations, rather than in fluctuations of the bosonic order parameter. The instability towards incommensurate 
order is associated with particular deformations of the Fermi surface. These alter the spectrum of the electronic soft modes which couple to the magnetic 
order parameter, leading to a self-consistent lowering of the free energy. The spiral is more stable than the homogeneous ferromagnet because the corresponding 
elliptical Fermi-surface deformations increase the surface-to-volume ratio and therefore enlarge the phase space available for low energy, 
particle-hole modes.

Fluctuation-driven spiral phases near ferromagnetic quantum critical points are within the realm of experimental detection. The first clear example is PrPtAl, 
where detection by neutron diffraction is possible because of an amplification of magnetic moments due to the coupling between conduction electrons 
and local spins.\cite{Jabbar+13}  In the presence of disorder, long-range order in the spiral phase is destroyed, leading to a helical glass with a 
highly anisotropic correlation length.\cite{franksteven} Such a glassy state appears to have been observed recently near the avoided ferromagnet quantum critical point of  
CeFePO (Ref.~\onlinecite{brando}). Since fluctuation-driven first-order behavior is such a robust feature of itinerant ferromagnets,\cite{belitzkirkpatrick} we expect that 
the instability towards incommensurate order should be similarly generic and also occur in quasi two-dimensional systems.

In this paper, we employ the fermionic quantum order-by-disorder approach to demonstrate that the phase diagram in $d=2$ has the same topology as in $d=3$ and 
exhibits a spiral phase at low temperatures. We perform a re-summation of the leading divergences, which enables us to follow the phase boundaries of the 
spiral phase far from the tricritical point down to $T=0$. The analytic structure depends crucially on the spatial dimension. In $d=3$, the re-summation of the leading divergent terms yields 
a contribution to the free energy $\Delta F_1 \sim M^4 \ln (M^2 +T^2)$, in agreement with previous results \cite{Belitz+99}. In $d=2$ we find 
$\Delta F_1 \sim M^2 T$ and $\Delta F_1 \sim M^3 \ln M$, in the regimes  $T \gg M$ and $T \ll M$ respectively. These expressions 
are generalized for the modulated spiral state. We show that the low-temperature behavior is controlled by a hierarchy of divergences and we derive a recursion 
relation that relates all sub-leading divergences to the leading term $\Delta F_1$. We use this recursion to re-sum all corrections at $T=0$ to find the exact location of the quantum critical point in $d=3$ from the expression $\Delta F \sim M^4 \ln (M)$.

The outline of this paper is as follows. In Sec.~\ref{sec.model} we introduce our model and summarize the key steps of the field-theoretical derivation of the 
free energy, which is written as a functional of the electron dispersion in the presence of spiral ferromagnet order. We further illustrate how the free energy of the incommensurate 
state can be deduced from the expression for the homogeneous state. In Sec.~\ref{sec.resum} we analyze the fluctuation integral and classify the contributions 
that diverge as $T \rightarrow 0$ in terms of the number of derivative operators they contain. We obtain closed-form expressions in $d=2$ and $d=3$
from a re-summation of the leading divergences of all orders. To go beyond this, we derive a recursion relation for the sub-leading terms, and do the full re-summation for $T=0$ in $d=3$. The resulting phase diagrams are presented 
in Sec.~\ref{sec.phase} and our findings are discussed in Sec.~\ref{sec.disc} in the context of other work and possible extensions.

\section{Quantum order-by-disorder framework}
\label{sec.model}

\subsection{Electronic Hamiltonian}

We work with a model of itinerant electrons in $d$-dimensions at chemical potential $\mu$. We allow the electrons to interact via a contact Hubbard repulsion of strength $g$. 
The Hamiltonian of this model is
\be
\mathcal{H} = \underset{{\bf{k}},\sigma}{\sum} (\epsilon_{\bf{k}} - \mu) \hat{n}_{{\bf{k}},\sigma} + g \int \, d^d {\bf{r}} \, \hat{n}_\uparrow ({\bf{r}}) \hat{n}_\downarrow ({\bf{r}}),
\label{Ham}
\ee
where we measure momenta in units of the Fermi momentum $k_F$. We only consider an isotropic free electron dispersion $\epsilon_{\bf{k}} = k^2/2$,
which does not allow for a nesting of the Fermi surface. While the mean-field theory of this model does not predict any incommensurate  states, fluctuations have 
been shown to stabilize a modulated spiral state close to the underlying ferromagnetic quantum critical point in $d=3$. In the remainder of this Section, we will briefly 
revisit the fermionic quantum order-by-disorder approach, keeping the spatial dimension general. The different non-analytic structures of the fluctuations in $d=2$ and $d=3$ will be analyzed in detail in Sec.~\ref{sec.resum}.

\subsection{Free-energy functional}

The key steps in deriving the free energy functional  are as follows. i. Starting from a coherent state path integral, we perform a 
Hubbard-Stratonovich decoupling of the electron interaction term in spin- and charge channels. ii. We decompose the fluctuation fields introduced in this way into zero- and 
finite-frequency components. The former correspond to static order in the system. Here, we include the spiral magnetic order parameter 
\be
\mathbf{M}_Q(\mathbf{r}) = M [ {\bf{n}}_x \cos (\bf{Q} \cdot \bf{r} ) +{\bf{n}}_y \sin (\bf{Q} \cdot \bf{r} )] 
\ee
with ${\bf{Q}} = Q {\bf{n}}_z$ in the free-fermion propagator. This facilitates the self-consistent expansion and  re-sums particular classes of diagrams to infinite order. 
iii. We trace over the fermions, keeping all terms up to quadratic order in the finite-frequency fluctuation fields. iv. We perform the Gaussian integrals over the 
fluctuation fields. v. Finally, we do the summation over Matsubara frequencies.

This procedure yields a general expression for the free energy $\mathcal{F}(M,Q)$ as a functional of the electron 
mean-field dispersion $\epsilon_{\bf{k}}^\sigma$ in the presence of spiral order,
\be
\epsilon_{\bf{k}}^\pm = \frac{k^2}{2} \pm \sqrt{({\bf{k}} \cdot {\bf{Q}} )^2 + (gM)^2}.
\label{mfdisp}
\ee
At mean-field level we obtain
\be
{\mathcal{F}}_{\textrm{mf}} (M,Q) = g M^2-T \underset{{\bf{k}},\sigma = \pm}{\sum} \ln [1+ e^{-(\epsilon_{\bf{k}}^{\sigma} - \mu )/T}],
\label{freeenergy}
\ee
while the fluctuation contribution is given by the integral 
\be
\Delta \mathcal{F} (M,Q) = 2g^2 \overset{\prime}{\underset{ {\bf{k}}_1, \dots, {\bf{k}}_4}{\sum}} \frac{ n({\epsilon}_{{\bf{k}}_1}^+) n({\epsilon}_{{\bf{k}}_2}^-) 
[n({\epsilon}_{{\bf{k}}_3}^+) +n({\epsilon}_{{\bf{k}}_4}^-)]}{\epsilon_{{\bf{k}}_1}^+ +\epsilon_{{\bf{k}}_2}^- -\epsilon_{{\bf{k}}_3}^+ -\epsilon_{{\bf{k}}_4}^-}.
\label{freefluct}
\ee
The summation runs over the momenta ${\bf{k}}_1,\dots,{\bf{k}}_4$ subject to the constraint ${\bf{k}}_1+{\bf{k}}_2 = {\bf{k}}_3+{\bf{k}}_4$ and 
$n(\epsilon) = 1/[1 + e^{(\epsilon - \mu)/T}]$ denotes the Fermi function. Note that this result is derived from self-consistent second-order perturbation theory
after subtraction of an unphysical UV divergence.\cite{una,pathria}

\subsection{Angular averages}

The free energy is a functional of the electron dispersion Eq.~(\ref{mfdisp}). As a result of this, $M$ and $Q$ enter the free energy in a similar manner, and since derivatives of the integrands 
in Eq.~(\ref{freeenergy}) and Eq.~(\ref{freefluct}) are strongly peaked near $k_F$, there exist simple proportionalities between the finite-$Q$ and $Q=0$ coefficients.
The proportionality factors are determined by combinatorial factors and angular averages over powers of 
\be
\eta_{\mathbf{k}}^2 =\frac{(\mathbf{k\cdot Q})^2}{k_F^2 Q^2}.
\ee
For example, the ratio of 
the $Q^2M^2$ and $M^4$ coefficients is given by $2\langle \eta_{\mathbf{k}}^2\rangle$. The fact that the two coefficients become negative at the same time explains why the 
fluctuation-driven first-order transition to the homogeneous ferromagnet is pre-empted by a transition into a spiral state. 

The angular averages depend on the dimension $d$ and are defined as  $\langle \dots \rangle = \Gamma_{d}^{-1} \int \, d \Omega_d \dots$ with $\Omega_d$ the angular 
part of the volume element  and $\Gamma_{d}$ the surface area of a unit-sphere in $d$-dimensions. The relevant averages are easily calculated as 

\be
\langle \eta_{\mathbf{k}}^{2n}\rangle =  \begin{cases}  (2n-1)!!/(2n)!! & (d =2) \\
       1/(2n+1) &(d =3) \end{cases},
\ee
where $!!$ denotes the double factorial function.\cite{doubfac} These observations enable us to calculate the free energy $\mathcal{F} (M,Q)$ of the spiral state from the free energy $F(M):=\mathcal{F} (M,Q=0)$ for the homogeneous 
ferromagnet by the equation   
\be
\mathcal{F}(M,Q) = \left\langle F\left(\sqrt{M^2 + \eta_{\mathbf{k}}^2 Q^2}\right) \right\rangle - \langle F(\eta_{\mathbf{k}} Q) \rangle ,
\label{qspiral}
\ee
where we have rescaled $Q$ by $g$ for convenience. Note that the free energy only contains even powers of $M$ and $Q$ since the electronic 
Hamiltonian (\ref{Ham}) is isotropic and does not break inversion symmetry, $\mathbf{r}\to\mathbf{-r}$.

\subsection{Mean-Field contribution}

The coefficients $\alpha_{\textrm{mf}}$, $\beta_{\textrm{mf}}$, and $\gamma_{\textrm{mf}}$
in the Landau expansion for the homogeneous ferromagnet,
\be
F_{\textrm{mf}}(M) = \alpha_{\textrm{mf}} M^2 + \beta_{\textrm{mf}} M^4 + \gamma_{\textrm{mf}} M^6+\ldots, 
\label{meanfieldlandau}
\ee
are given by integrals over derivatives of the Fermi function,
\be
\begin{split}
\alpha_{\textrm{mf}} &= g +\frac{2g^2}{2!} \int \, d^d {\bf{k}} \, n^{(1)}(\epsilon_{\bf{k}}) , \\
\beta_{\textrm{mf}} &= \frac{2g^4}{4!} \int \, d^d {\bf{k}} \, n^{(3)}(\epsilon_{\bf{k}}) , \\
\gamma_{\textrm{mf}} &= \frac{2g^6}{6!} \int \, d^d {\bf{k}} \, n^{(5)}(\epsilon_{\bf{k}}). 
\label{mfints}
\end{split}
\ee
For the isotropic dispersion $\epsilon_{\bf{k}}=k^2/2$, these reduce to simple one-dimensional integrals. Using Eq.~(\ref{qspiral}), we obtain the mean-field 
contribution
\bea
\mathcal{F}_{\textrm{mf}}(M,Q) & = &  \left(\alpha_{\textrm{mf}} +2\beta_{\textrm{mf}}\langle \eta_{\mathbf{k}}^2\rangle Q^2+3\gamma_{\textrm{mf}}\langle 
\eta_{\mathbf{k}}^4 \rangle Q^4\right)M^2 \nonumber \\ 
& & + \left(\beta_{\textrm{mf}} +3 \langle \eta_{\mathbf{k}}^2\rangle Q^2 \right) M^4 + \gamma_{\textrm{mf}} M^6,
\eea
to the free energy of the spiral ferromagnet.

\section{Resummation of fluctuations}
\label{sec.resum}

In this section we analyze the analytic structure of the fluctuation integral $\Delta \mathcal{F} (M,Q)$ Eq.~(\ref{freefluct}) in general dimension 
to obtain closed-form expressions from a re-summation of the leading divergences in $d=3$ and $d=2$. We first focus on the fluctuation 
integral of the homogeneous ferromagnet,
$\Delta F(M) =\Delta \mathcal{F} (M,Q=0)$, which is given by Eq.~(\ref{freefluct}) with $\epsilon_{\bf{k}}^{\sigma} = k^2/2 - \sigma g M$, and then generalize to finite $Q$,  
using Eq.~(\ref{qspiral}). 

It is most convenient to express $\Delta F(M)$ in terms of the components of the particle-hole density of states 
$\rho(q,\epsilon) = \tilde{\rho} (q,\epsilon) - \Delta \rho (q,\epsilon)$, where these are defined as 
\be
\begin{split}
{\tilde{\rho}}^{\sigma}(q,\epsilon) &= \int \, d^d {\bf{k}} \, n( \epsilon^\sigma_{{\bf{k}}-\frac{{\bf{q}}}{2}}) \delta (\epsilon -  \epsilon^\sigma_{{\bf{k}}+\frac{{\bf{q}}}{2}}+ \epsilon^\sigma_{{\bf{k}}-\frac{{\bf{q}}}{2}}), \\
\Delta \rho^{\sigma}(q,\epsilon) &= \int \, d^d {\bf{k}} \, n( \epsilon^\sigma_{{\bf{k}}-\frac{{\bf{q}}}{2}}) n( \epsilon^\sigma_{{\bf{k}}+\frac{{\bf{q}}}{2}}) \delta (\epsilon -  \epsilon^\sigma_{{\bf{k}}+\frac{{\bf{q}}}{2}}+ \epsilon^\sigma_{{\bf{k}}-\frac{{\bf{q}}}{2}}),
\end{split}
\label{phDOS}
\ee
and calculated in Appendix A for dimension $d=3$ and $d=2$. With these definitions of $\tilde{\rho}$ and $\Delta \rho$, the fluctuation contribution to the free energy 
becomes
\be
\Delta F (M) = 2g^2 \underset{\sigma = \pm}{\sum} \int_{{\bf{q}},\epsilon_1,\epsilon_2} \, \frac{\Delta \rho^{\sigma}({\bf{q}},\epsilon_1) {\tilde{\rho}}^{-\sigma}(-{\bf{q}},\epsilon_2)}{\epsilon_1 + \epsilon_2}.
\label{fluctint}
\ee
Here the integrals run over ${\bf{q}} \in {\mathbb{R}}^d$ and $\epsilon_1 , \epsilon_2 \in \mathbb{R}$.

\subsection{Landau Expansion of $\Delta F$}

We generate a Landau expansion of the fluctuation contributions for $Q=0$ by Taylor expanding the expression Eq. (\ref{fluctint}) for small $M$, 
\be 
\Delta F(M) = \alpha_{\textrm{fl}} M^2 + \beta_{\textrm{fl}} M^4 + \gamma_{\textrm{fl}} M^6 + \dots ,
\ee
where e.g. $\alpha_{\textrm{fl}} = \partial_{M^2} \Delta F|_{M=0}$.  We exchange derivatives with respect to $M$ for derivatives with respect to $\mu$ using the relation $ \sigma g \partial_\mu = -\partial_M$ to write these coefficients in terms of integrals of the form 
\be 
J_{m,n} (T)=  \int_{{\bf{q}},\epsilon_1,\epsilon_2} \, \frac{\Delta {\rho}^{(m)}({\bf{q}},\epsilon_1) \tilde{\rho}^{(n)}(-{\bf{q}},\epsilon_2)}{\epsilon_1 + \epsilon_2}.
\label{jdefn}
\ee

These integrals depend only on derivatives of the particle-hole density of states with respect to $\mu$  for $M=0$. We use the shorthand notation 
${\tilde{\rho}}^{(m)} = \partial_\mu^m \tilde{\rho}$ and $\Delta \rho^{(m)} = \partial_\mu^m \Delta\rho$. Using this notation, we can write the $M^2$ coefficient as

\be
\begin{split}
\alpha_{\textrm{fl}} &= \frac{4g^4}{2!}( J_{0,2} -2J_{1,1}+J_{2,0}) \\ 
&= \frac{g^2}{2!} (\partial_{\mu}^2 \Delta F|_{M=0} - 16 g^2 J_{1,1}).
\end{split}
\ee

In the second line we have used integration by parts to write the coefficient in terms of the symmetric integrals $J_{1,1}$ and $ \Delta F|_{M=0} = 4g^2 J_{0,0}$.
Repeating this process up to order $M^{10}$, we find the coefficients listed in Table~\ref{Table1}. In the limit $T\to 0$, the integrals $J_{n,n}(T)$ 
diverge for $n\ge2$ in $d=2,3$. The divergence becomes stronger with increasing order $n$ and therefore  as we move from 
left to the right in each row of Table~\ref{Table1}. At each order $M^{2n}$ in the Landau expansion, the  leading small-$T$ dependence comes from the term 
proportional to $J_{n,n}(T)$.

\begin{center}
\begin{table*}[t]
{\small
\hfill{}
\renewcommand{\arraystretch}{1.5}
\begin{tabular}{  c | | c  c  c  c  c  c }
$\alpha_{\textrm{fl}}$  & $\frac{g^2}{2!} \partial_{\mu}^2 \Delta F|_{M=0}$ & $-\frac{16g^4}{2!} J_{1,1}$ & & & & \\ \hline
$\beta_{\textrm{fl}}$  & $\frac{g^4}{4!} \partial_{\mu}^4 \Delta F|_{M=0}$ & $-\frac{32g^6}{4!}\partial_{\mu}^2 J_{1,1}$ & $+\frac{64g^6}{4!} J_{2,2}$ & & & \\ \hline
 $\gamma_{\textrm{fl}}$  & $\frac{g^6}{6!} \partial_{\mu}^6 \Delta F|_{M=0}$ & $-\frac{48g^8}{6!} \partial_{\mu}^4 J_{1,1}$ & $+\frac{192 g^8}{6!} \partial_{\mu}^2 J_{2,2}$ & $-\frac{256 g^8}{6!} J_{3,3}$ & & \\ \hline
$\delta_{\textrm{fl}}$  & $\frac{g^8}{8!} \partial_{\mu}^8 \Delta F|_{M=0}$  & $-\frac{64 g^{10}}{8!} \partial_{\mu}^6 J_{1,1}$ & $+ \frac{384 g^{10}}{8!} \partial_{\mu}^4 J_{2,2}$ & $ -\frac{1024 g^{10}}{8!} \partial_{\mu}^2 J_{3,3}$ & $+\frac{1024 g^{10}}{8!} J_{4,4}$ & \\ \hline
$\eta_{\textrm{fl}}$  & $\frac{g^{10}}{10!} \partial_{\mu}^{10} \Delta F|_{M=0}$ & $ -\frac{80 g^{12}}{10!} \partial_{\mu}^8 J_{1,1}$ & $ +\frac{640 g^{12}}{10!} \partial_{\mu}^6 J_{2,2}$ & $ -\frac{2560 g^{12}}{10!} \partial_{\mu}^4 J_{3,3}$ & $+\frac{5120 g^{12}}{10!} \partial_{\mu}^2 J_{4,4}$ & $-\frac{4096 g^{12}}{10!} J_{5,5}$ \\ 
  \vdots & \vdots & \vdots & \vdots & \vdots & \vdots & \vdots \\
\end{tabular}}
\hfill{}
\caption{Table of the dependency of expansion coefficients in the Landau expansion for the homogeneous ferromagnet on the integrals $J_{n,n}$.}
\label{Table1}
\end{table*}
\end{center}

The behavior of phase boundaries very close to the finite-temperature tricritical point is controlled by the smallness of $M$ and therefore by the coefficients 
$\alpha$, $\beta$, and $\gamma$. The phase boundaries at low temperatures far from the tricritical point are determined by the leading divergences as 
$T\to0$. Using Table~\ref{Table1}, we may begin to spot patterns in the coefficients of $J_{n,n}$. On the leading diagonal of the table, we note that 
the coefficient of $J_{n,n}$ takes a particularly simple form of ${(-1)^n 2^{2n}}/{(2n)!}$. Re-summing the terms of all orders in $M$ along this diagonal we obtain
\be 
\Delta F_1 (M) = 4g^2 \overset{\infty}{\underset{n=1}{\sum}} \frac{(-1)^n (2gM)^{2n}}{(2n)!} J_{nn}(T).
\label{deltaF1}
\ee

We later calculate closed-form expressions for $\Delta F_1$  in $d=3$ and $d=2$. Continuing down the chain to the $m$th diagonal, we obtain the  
free energy contribution
\be 
\begin{split}
&\Delta F_m (M) = g^{2m} {\partial}_{\mu}^{2m} 4g^2 \\
& \,\,\,\,\, \times
\overset{\infty}{\underset{n=1}{\sum}} \frac{(-1)^n (2g)^{2n} (2n-1)!! M^{2(m+n)}}{(2n)!  (2n+2m-1)!! m! \, 2^n} J_{nn}(T) .
\label{deltaferromagnet}
\end{split}
\ee
Using this expression, we find a differential equation which relates $\Delta F_m$ to $\Delta F_{m-1}$,
\be
\partial_M \left(\frac{{\Delta F}_{m}}{M} \right) = \frac{ g^2 \partial_\mu^2  \Delta F_{m-1}}{2m}.
\label{recursion}
\ee 
As we approach $T=0$, we find that sub-leading corrections become more significant, and so to find the correct quantum critical point, we must be able to calculate them in this limit. Repeated application of Eq.~(\ref{recursion}) enables us to calculate all sub-leading corrections from the functional form of the leading re-summed correction 
$\Delta F_1$. Re-summation of all these sub-leading corrections is a tractable calculation in $d=3$.

\subsection{Calculation of $\Delta \mathcal{F}_1$ and $\Delta \mathcal{F}$ in $d=3$}

We proceed by deriving an explicit expression for the re-summation $\Delta F_1(M)$ (\ref{deltaF1}) of leading divergences in $d=3$. Details of the calculation 
of particle-hole density of states (\ref{phDOS}) and the integrals $J_{n,n}(T)$ (\ref{jdefn}) are given in Appendices A and B, respectively. After summation over $n$, 
the final result takes the form
\be 
\Delta F_1 (M) = -2\lambda(1+\ln2)M^2+\lambda M^4 \ln\left(\frac{4g^2 M^2 + T^2}{4\mu^2}\right), 
\label{leading3d}
\ee
where we have defined $\lambda=4g^6 \nu_F^3/ (3 \mu^2)$ with $\nu_F = {k_F}/{2 \pi^2}$ and $\mu = {k_F^2}/{2}$. Note that we have included the non-divergent
fluctuation correction of order $M^2$, corresponding to the $n=1$ term in the sum. As noted in previous work \cite{una}, this term changes the location of the tricritical point and the phase behavior near to
it. The logarithmic contribution arrises from the summation of all divergent terms $n\ge 2$ and is 
of the same form as the diagrammatic result.\cite{Belitz+99}

Using this re-summed form and the relation (\ref{qspiral}), we can also obtain a closed-form expression for the leading fluctuation correction 
$\Delta {\mathcal{F}}_1(M,Q)$ to the free energy of the spiral ferromagnet, 
\be 
\begin{split}
&\Delta {\mathcal{F}}_1(M,Q)/\lambda=-2(1+\ln2)M^2\\
& +\left(M^4 +\frac{2}{3} M^2 Q^2 + \frac{1}{5} Q^4\right)\ln \left[\frac{4g^2 (M^2+Q^2) + T^2}{4\mu^2} \right] \\ 
& - \frac{1}{5} Q^4 \ln \left(\frac{4g^2 Q^2 + T^2}{4\mu^2} \right) - \frac{14}{45} M^2 Q^2\\
&  +\left[ \frac{16}{15} M^4 -\frac{8}{15} M^2\frac{T^2}{4g^2} + \frac{2}{5} \biggl( \frac{T^2}{4g^2} \biggr)^2 \right] \\
&  \times\left[ \frac{\sqrt{M^2 +\frac{T^2}{4g^2}}}{Q} \arctan \left( \frac{Q}{\sqrt{M^2 + \frac{T^2}{4g^2}}} \right)-1 \right].
\end{split}
\label{spiral3d}
\ee

Given the form of $\Delta F_1$ in Eqn.~(\ref{leading3d}), we may then use the recursion relation (\ref{recursion}) to calculate the full, all-orders quantum correction $\Delta F(M) = \sum_{n=1}^\infty \Delta F_n (M,0)$ at $T=0$ (see Appendix C). The result is that 

\be 
\Delta F (M) = 2\sqrt{\pi} \lambda M^4 \ln \biggl[ \frac{g^2 M^2}{\mu^2} \biggr].
\ee

Having calculated the form of the leading correction to the free energy for general $T$ and $M$, $\Delta F_1$, and the re-summation of all the divergent contributions to the free energy at $T=0$, $\Delta F$, we may now postulate an analytical form for the full, re-summed quantum fluctuation contribution to the free energy that interpolates between these two cases. We suggest the functional form \cite{alpha}

\be 
\Delta F (M) = \lambda M^4 \ln \biggl[ \biggl(\frac{4 g^2 M^2}{\mu^2} \biggr)^{2\sqrt{\pi}} + \frac{T^2}{\mu^2} \biggr].
\label{fullresum}
\ee

This expression encapsulates the leading corrections in the vicinity of the tri-critical point, where $\Delta F \sim M^4 \ln (T)$, and also gets the precise location of the $T=0$ intercept correct, thereby capturing the important physics in a simple, closed form.

\subsection{Calculation of $\Delta \mathcal{F}_1$ in $d=2$}

In $d=2$, it has not proved possible to write down a closed, analytic form for the first re-summed correction to the free energy, $\Delta F_1$. 
However, it is still possible to use our formalism to find the asymptotic forms of this correction in the limits $T \ll M$ and $T \gg M$.
These are given by
\be 
\Delta  F_1 (M) = \begin{cases} c_- M^3 \ln M & \text{for} \, T \ll M \\c_+ M^2 T & \text{for} \, T \gg M  \end{cases} 
\ee
with coefficients $c_-=16g^4 \pi^{3/2}/[(2\pi)^5 \Gamma (3/2)]$ and $c_+=16\sqrt{2} g^4 \pi^{3/2} \text{Li}_{1/2}(-1)/(2\pi)^5$. Details of the derivation are 
given in Appendix D.

Again, we perform the angular averages (\ref{qspiral}) to generalize to finite $Q$ and investigate possible instabilities towards spiral formation. 
The high-temperature form modifies the mean field coefficient of $M^2$, adding a term linear in $T$, which cannot contribute to a term $\sim Q^2$.
In the regime of low temperatures, $T\ll M$, the fluctuation corrections generate extra terms in the Landau expansion of $\mathcal{F}(M,Q)$,
\bea
\Delta \mathcal{F}_1 (M,Q) & = &  c_- \left[ M^3 \ln M + \frac 12 Q^2 M\left( 1+ 3 \ln M \right)\right. \nn\\
& & + \left.\frac{3}{16}Q^4 M^{-1}\left( 1 + \frac{3}{4} \ln M \right) \right],
\label{spiral2d}
\eea
where we have included terms up to order $Q^4$.

\section{Phase diagrams}
\label{sec.phase}

We are now in a position to calculate the phase diagrams in $d=3$ and $d=2$ by minimizing of the free energy $\mathcal{F}(M,Q) =\mathcal{F}_\textrm{mf}(M,Q)
+\Delta \mathcal{F} (M,Q)$. In $d=3$, we use the proposed interpolating form (\ref{fullresum}), whereas in $d=2$, we approximate $\Delta \mathcal{F}(M,Q)$ by the leading re-summed correction $\Delta \mathcal{F}_1 (M,Q)$.  We show that despite the different analytic structures of the fluctuation 
contributions contained in $\Delta \mathcal{F}$, the phase diagrams in two and three dimensions have the same topology, and both show an instability to a spiral state below the temperature 
$T_c$ of the tricrital point. The spiral phase intervenes between the homogeneous ferromagnet at strong electron repulsions $g$ and the paramagnet at small $g$. 
Let us start by writing conditions for the different phase transitions we seek. Note that the free energy is defined such that $\mathcal{F}(M=0,Q)=0$.

\begin{itemize}
\item[i.] The second-order transition from the paramagnet into the ferromagnet for $T>T_c$ as the coupling $g$ is increased is obtained by $\partial_{M^2}F|_{M=0} = 0$. 
(As before we have defined $F(M)=\mathcal{F}(M,Q=0)$.)

\item [ii.] A first-order transition between the paramagnet and the ferromagnet as $g$ is increased when $T<T_c$ occurs when $F(M^\star) =0$ and 
$\partial_{M^2}F|_{M=M^\star}=0$ for some $M^\star \neq 0$. 

\item[iii.] Allowing for the possibility of spiral states, we find that the first-order transition from paramagnet into ferromagnet is pre-empted by a first-order transition into a spiral phase. This is slightly more involved; 
we must first solve $\partial_{Q^2}\mathcal{F}(M,Q) = 0$ to obtain the optimal pitch $\tilde{Q}(M)$ for a given magnetization $M\neq 0$. We then look for a first-order 
transition in $\tilde{F}(M)=\mathcal{F}(M,\tilde{Q}(M))$ as described in (ii).

\item[iv.] A Lifshitz transition from the spiral state into the ferromagnet, where the pitch $Q$ of the spiral goes smoothly to zero. This is given by the line along which 
$\tilde{Q}= 0$.
\end{itemize}

\subsection{Phase Diagram in $d=3$.}

A phase diagram in $d=3$ has been calculated in Ref.~\onlinecite{una}, taking into account fluctuation corrections to the $M^2$, $M^4$, and $M^2 Q^2$ coefficients.  
In this section we will show that while this approximation is valid in the vicinity of the tricritical point, it fails to describe the behavior at low temperatures. 
Since the order of the divergences of the coefficients increases with the order in the Landau expansion, it is crucial to use the re-summation (\ref{spiral3d}) of leading 
divergences to obtain the phase boundaries at low temperatures.

For comparison, we recalculate the phase diagram of Ref.~\onlinecite{una} from the truncated Landau expansion 
\bea
\mathcal{F}_\textrm{trunc}(M,Q) & = & \left(\alpha+\frac23 \beta Q^2+\frac 35 \gamma Q^4\right)M^2 \nn\\
& & +(\beta+\gamma Q^2)M^4+\gamma M^6,
\eea
including fluctuation corrections $\alpha_\textrm{fl}=-2\lambda(1+\ln2)$ and 
$\beta_\textrm{fl}=2\lambda (1+\ln T)$ up to quartic order. Note that the $\ln T$-contribution of $\beta_\textrm{fl}$ arises from the zeroth order term of an expansion 
of the logarithm in $\Delta F_1$ (\ref{leading3d}) while the non-divergent contribution to $\beta_\textrm{fl}$ comes from the sub-leading terms in the second
row of Table~\ref{Table1}.

The second-order phase boundary between the ferromagnet and the paramagnet is determined by $\alpha=0$, consistent with condition (i). Fluctuations render the transition 
first-order below the temperature $T_c$ of the tricritical point which is determined by the intersection of the $\alpha=0$ and $\beta=0$ lines. For contact repulsion, 
we obtain $T_c/\mu\approx 0.3$, in agreement with previous work.\cite{belitz1,andrewgareth,una} 

The tricritical temperature is considerably reduced by disorder\cite{belitz1}
and finite range interactions.\cite{belitz1,conduitfinite} With increasing range of electron interactions, the relative strength of fluctuation corrections $\lambda$ decreases, 
leading to an exponential suppression of $T_c$. This is apparent from the condition $\beta=\beta_\textrm{mf}+2\lambda (1+\ln T_c)=0$. 

In order to determine the first-order 
transition line between the spiral and the paramagnet, we follow the recipe described under (iii). This leads to the condition $\alpha \gamma = \frac{17}{63} \beta^2$
for the phase boundary.\cite{una} Finally, the condition $\tilde{Q} =0$ which defines the Lifshitz line between the spiral and the ferromagnet (iv) coincides
with the $\alpha = 0$ line below $T_c$.

\begin{figure}[t!]
  \centering
    \includegraphics[width=0.9\linewidth]{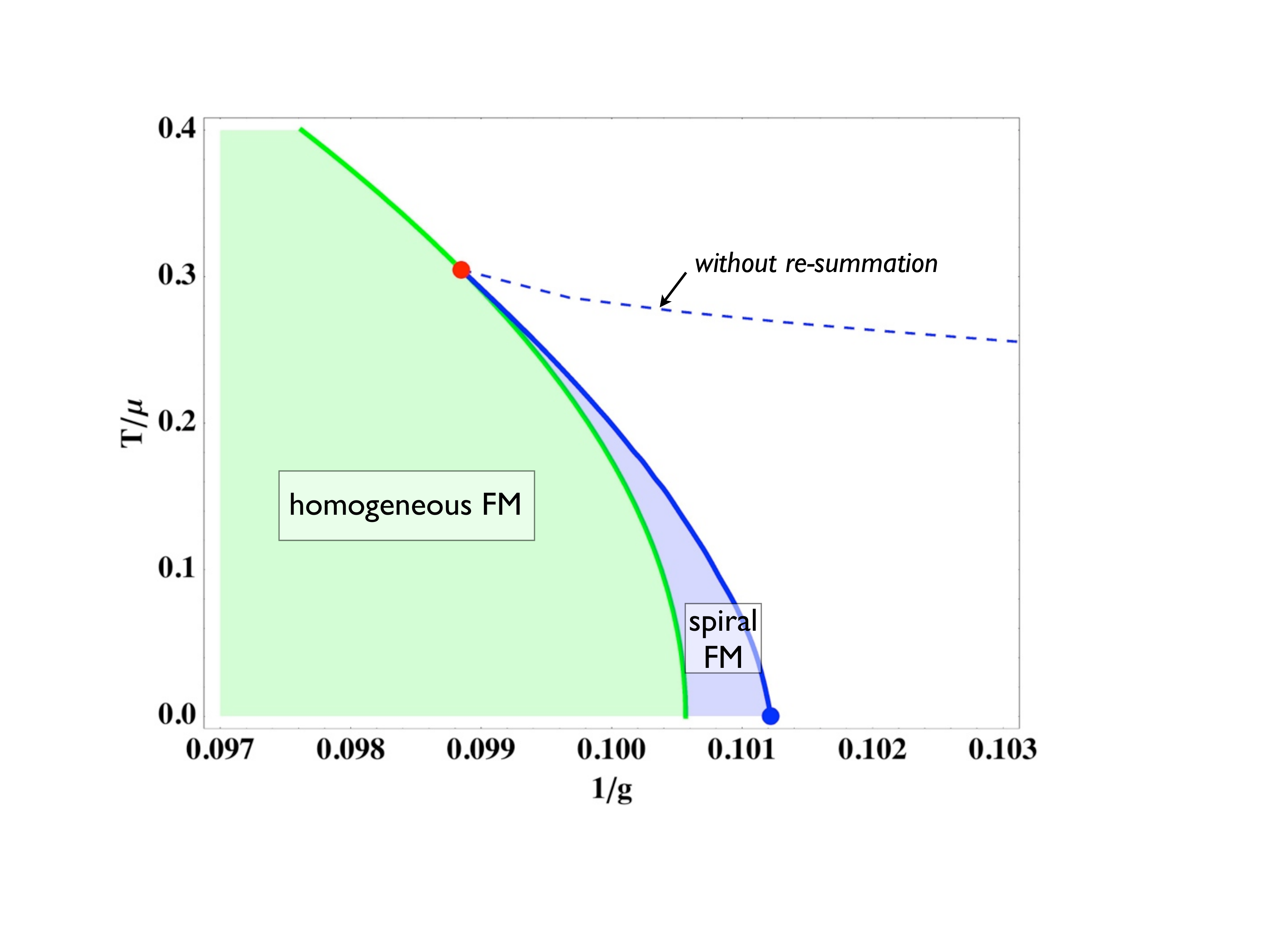}
  \caption{(color online) Phase diagram in $d=3$ as a function of inverse electron repulsion $1/g$ and temperature $T/\mu$. Above the tricritcal point (shown in 
  red) we find a continuous transition between the ferromagnet at large $g$ and a paramagnet at small $g$. The spiral forms below the tricritical point between the paramagnet and
  the ferromagnet. The effect of the re-summation of leading divergencies is illustrated. (i) The dashed line shows the case without re-summation, which shows unphysical behavior as $T\to0$ (ii) Using the re-summation of the leading divergences $\Delta F_1$, the spiral region becomes invisible on the scale of this phase diagram, collapsing into a region very close to the homogeneous magnetic state. (iii) The solid line shows the phase boundary of the spiral found using the form for $\Delta F$ in Eqn.~(\ref{fullresum}), and gives the exact $T=0$ intercept, and the correct behaviour in the vicinity of the tricritical point.}
\label{resummedQ}
\end{figure}

The resulting phase diagram is shown in Fig.~\ref{resummedQ}. The low-temperature behavior of the first-order transition line between the paramagnet and the spiral 
phase is clearly unphysical. The phase boundary does not terminate on the $T=0$ axis but instead approaches zero temperature asymptotically as $g\to 0$, 
suggesting that there exists a transition into a spiral state at low temperatures for arbitrarily small values of $g$, far from the avoided quantum critical point.

In what follows, we will study the changes to the phase diagram which result from the inclusion of the functional form for $\Delta F_1 (M)$ given in Eqn.~(\ref{leading3d}) which gives just the leading re-summed corrections from quantum fluctuations, and the interpolating form $\Delta F(M)$ given in Eqn.~(\ref{fullresum}), which captures the divergences of all the higher-order terms in the Landau expansion. In order to investigate the spiral phase behaviour, we use the finite-$Q$ generalisation $\Delta F(M,Q)$ of Eqn.~(\ref{fullresum}), which is obtained by carrying out the angular averages.

The main effect of the 
re-summation is to cut off the $\ln T$-divergence by finite $M$.  This fundamentally changes the behavior of the first-order spiral/paramagnet line as illustrated 
in Fig.~\ref{resummedQ}. The phase boundary has a vertical intercept with the $T=0$ axis at a finite value $g_\kappa$, consistent 
with the Clausius-Clapeyron condition. Using just the leading re-summed expression $\Delta F_1 (M)$ (\ref{leading3d}), we are left with  
a very narrow region of spiral order. However, we know that for small values of $T$, the sub-leading re-summed corrections will become significant, so we must include all the subleading corrections too. We may explicitly calculate the location of the $T=0$ intercept for the first-order spiral transition line, to find a critical coupling $1/g_c \sim 0.1012$. Previous Monte Carlo analysis \cite{andrewgareth} suggests the transition into the spiral state at $T=0$ will occur at $1/g \approx 0.133$.  The numerical disagreement between the two approaches may stem from the fact that to carry out the $T=0$ Monte Carlo calculation, the contact interaction must be replaced by one with a negative finite range.

\subsection{Phase Diagram in $d=2$.}

\begin{figure}[t!]
  \centering
    \includegraphics[width=0.95\linewidth]{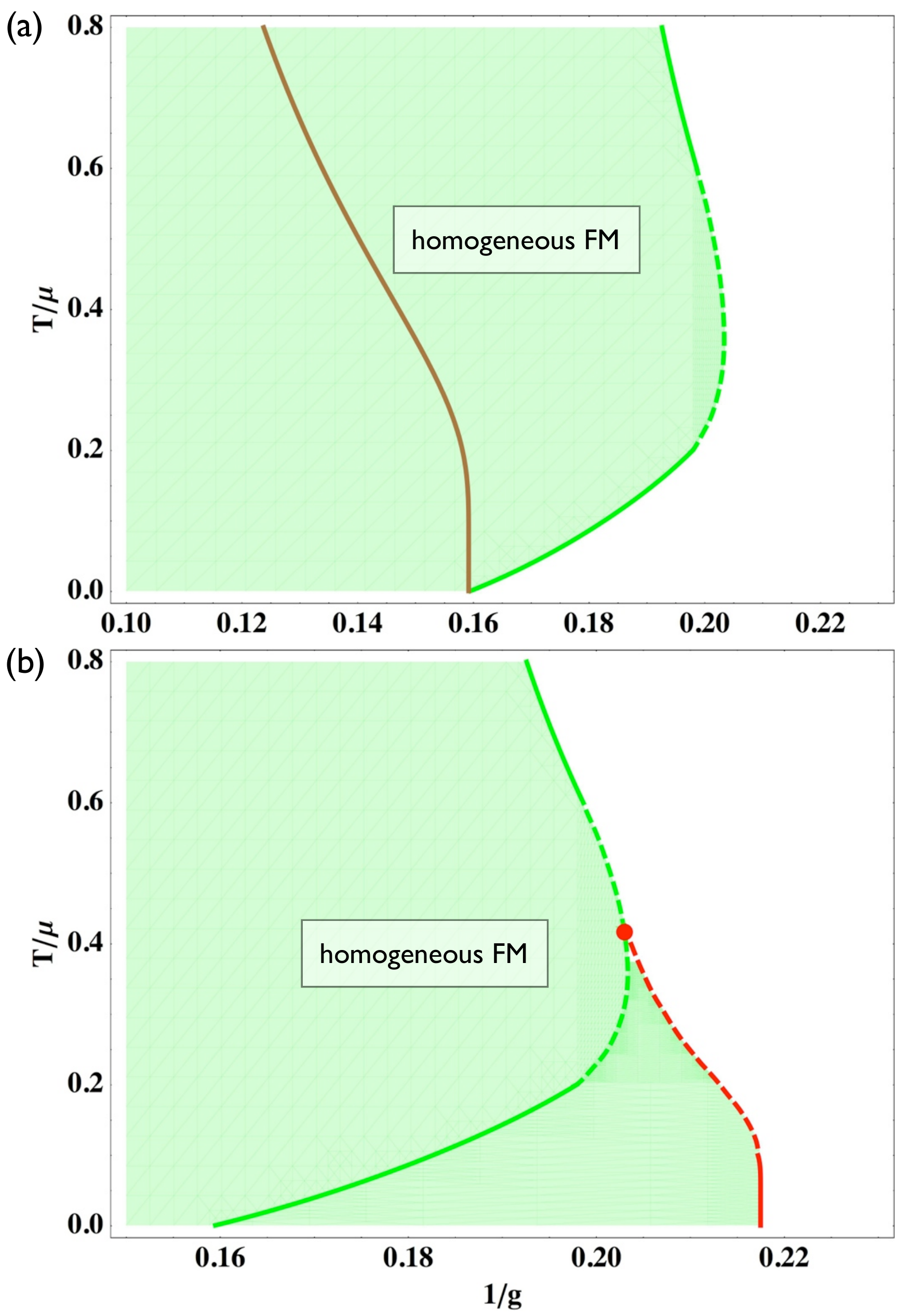}
  \caption{(color online) Intermediate phase diagrams in $d=2$ for $Q=0$. (a) Continuous phase boundaries between the ferromagnet and the 
  paramagnet obtained from the condition $\alpha = 0$, first using just the mean-field $\alpha =\alpha_{\textrm{mf}}$ (brown line), and then including the effects of fluctuations 
  $\alpha = \alpha_{\textrm{mf}} +\alpha_{\textrm{fl}}$ (green line).  In the regime $M\ll T\ll\mu$, $\alpha_{\textrm{fl}}=c_+ T$ ($c_+<0$), leading to a stabilization of ferromagnet 
  order. Around $T\simeq \mu$ fluctuations saturate, causing re-entrant behavior. We interpolate between the two regimes, indicated by a dashed green line.
  (b) The leading fluctuation correction $\Delta F_1(M)=c_- M^3 \ln M$ in the regime $T\ll M$ causes a first-order transition at low temperatures (solid red line). The 
  first-order line at higher temperatures is obtained by interpolation (dashed red line)  to the tip of the re-entrant $\alpha=0$ line, where one expects the location of the 
  tricritical point.}
\label{fig2and3}
\end{figure}

We proceed to calculate the phase diagram in $d=2$ by minimizing $\mathcal{F}(M,Q) =\mathcal{F}_\textrm{mf}(M,Q)
+\Delta \mathcal{F}_1(M,Q)$ with $\mathcal{F}_\textrm{mf}$ given by Eq.~(\ref{qspiral}) with $\langle \eta_{\mathbf{k}}^2\rangle =1/2$ and 
$\langle \eta_{\mathbf{k}}^4\rangle =3/8$ and $\Delta \mathcal{F}_1$ defined in Eq.~(\ref{spiral2d}). Note that since $\Delta \mathcal{F}_1$
is only known in  the regimes $T\ll M$ and $T\gg M$ [see  Eq.~(\ref{spiral2d})], we will only be able to determine the asymptotic behavior of 
the phase boundaries  and have to interpolate between the two regimes. There are crucial differences between the cases $d=2$ and 
$d=3$:

\begin{itemize}

\item  In two dimensions, the density of states is constant, leading to an 
exponentially weak temperature dependence of $\alpha_\textrm{mf}$. As a consequence, the critical interaction strength $g_c(T)$ for the  
mean-field transition of the ferromagnet is practically constant up to  $T\approx 0.2\mu$ [see Fig.~\ref{fig2and3}a]. 

\item The fluctuation contributions in $d=2$ are very different. 
Even in the vicinity of the tricritical point, we do not have a simple Landau expansion of the free energy; the corrections are intrinsically non-analytic 
across the whole phase diagram. 

\item The angular averages $\langle \eta_{\mathbf{k}}^{2n}\rangle$ are larger in $d=2$ and decay as
$1/\sqrt{n}$  for large $n$, opposed to $1/n$ in $d=3$. This has a profound effect on how the free energy of the spiral relates to that of the homogeneous ferromagnet. 

\end{itemize}

In spite of these differences, the phase diagram for the two-dimensional case turns out to be remarkably similar to that for three dimensions.
We proceed to construct the phase diagram in three steps. i.  We first analyze the effects of the fluctuation corrections $\Delta F_1(M)$ on the 
continuous transition between the ferromagnet and the paramagnet. This is controlled by the asymptotic form of $\Delta F_1$ in the regime $T\gg M$ since $M$ 
vanishes continuously at the second-order transition. ii. We determine the fluctuation-driven first-order transition, using the low temperature 
asymptotic form of $\Delta F_1$, which is valid for $T\ll M$. We extrapolate this first-order line to higher temperatures and estimate the position of 
the tricritical point. iii. We obtain the first-order spiral-to-paramagnet transition by minimizing $\mathcal{F}(M,Q)$ in the low-temperature regime. iv. Finally, we find that similar to the $d=3$ case, the Lifshitz line along which $Q \rightarrow 0$ coincides with the line $\alpha =0$.
This line is again extrapolated up to the tricritical point.

\emph{i. Second-Order Line.} Using the asymptotic expression $\Delta F_1(M)=c_+ M^2 T$, valid for  $\mu \gg T \gg M$, we obtain the fluctuation corrected 
transition line between the paramagnet and the ferromagnet by the condition $\alpha=\alpha_\textrm{mf}+c_+ T=0$. Since $\alpha_\textrm{mf}$ is almost constant at low 
temperatures and since $c_+<0$, the region of stability of the ferromagnet \emph{increases} with temperature where the phase boundary is almost linear. This is shown in 
Fig.~\ref{fig2and3}a. At higher temperatures, the fluctuation effects saturate, and the second-order line then tracks the mean-field transition, albeit 
at lower values of $g$. We expect that the true second-order transition line interpolates between these two asymptotic limits, leading to the re-entrant behavior 
sketched in Fig.~\ref{fig2and3}a. 

\begin{figure}[t!]
  \centering
    \includegraphics[width=0.95\linewidth]{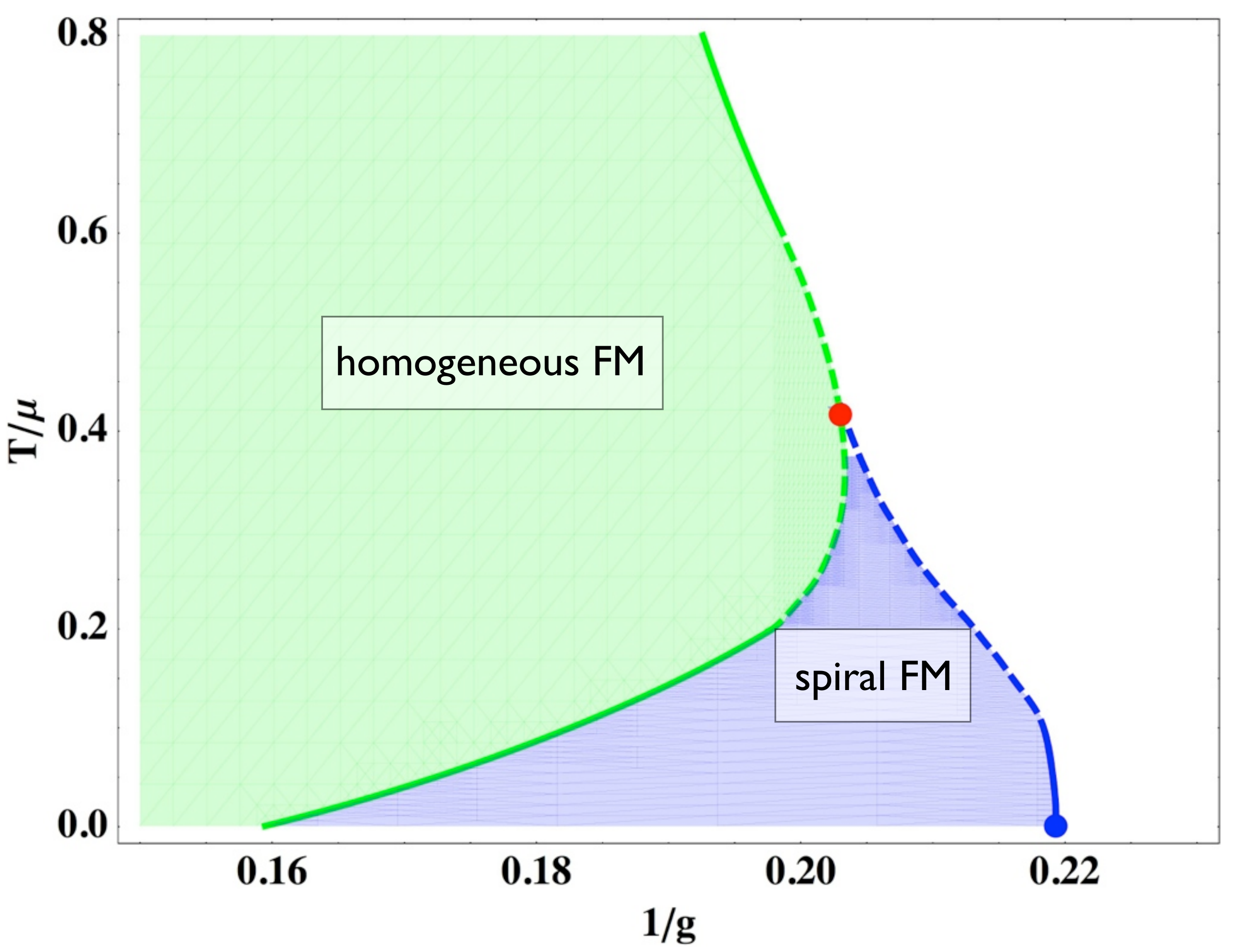}
  \caption{(color online) Phase diagram in $d=2$ as a function of inverse electron repulsion $1/g$ and temperature $T/\mu$. Dashed lines are interpolations
  between different asymptotic regimes. The phase diagram has the same topology as the one in $d=3$ (see Fig.~\ref{resummedQ}) and exhibits a spiral phase below the 
  tricritical point which is shown in red. As in $d=3$, the transition between the ferromagnet and the spiral is of the Lifshitz type while the spiral/paramagnet transition is first-order.}
\label{2dphase1st}
\end{figure}

\emph{ii. First-Order Line.} We now focus on the regime $T\ll M$ where the re-summed fluctuations are of the asymptotic form $\Delta F_1(M)=c_- M^3\ln M$
with $c_->0$. This contribution leads to a first-order transition between the ferromagnet and the paramagnet at low temperatures and values of $g$ that are considerably 
smaller than those determined by the condition $\alpha=0$ for the second-order line. To determine the first-order phase boundary, we numerically search for 
solutions $M\neq 0$ of the equations $F(M) =0$ and $\partial_{M^2}F=0$. If we follow this first-order line to higher $T$, we expect a tricritical point at the 
intersection with the $\alpha=0$. The most likely scenario is that this point coincides with the tip of the re-entrant $\alpha=0$ line, sketched in Fig.~\ref{fig2and3}b. 

\emph{iii. Spiral Phase.}  Finally, we allow for states where $Q\neq 0$. As found previously, the Lifshitz line where $Q \rightarrow 0$ is given by 
the extension of the second-order ferromagnet-to-paramagnet transition line to temperatures below the tricritical point.\cite{lifshitz}

We continue to calculate the first-order, paramagnet-to-spiral transition line in the limit $T\to 0$, using the asymptotic form of $\Delta\mathcal{F}_1(M,Q)$ in the 
regime $T\ll M$ (\ref{spiral2d}). We follow the same procedure we used in $d=3$ and first determine the optimal pitch $\tilde{Q}(M)$ for a given magnetization.
This requires an expansion of  $\mathcal{F}(M,Q)$ up to 4th order in $Q$. We then determine the first-order transition of $\tilde{F}(M)=\mathcal{F}(M,\tilde{Q}(M))$.
As in the $3d$ case we find that this first-order transition pre-empts the one into the homogeneous state. From the asymptotic form of the free energy, we can
only determine the phase boundary for small values of $T$ and rely on an interpolation up to the tricritical point. 
The resulting phase diagram is shown in Fig.~\ref{2dphase1st}.

\section{Conclusions and Discussion}
\label{sec.disc}

Despite its apparent simplicity, the Hamiltonian (\ref{Ham}) yields a rich and interesting phase diagram when we include the possibility of fluctuation-driven phases near to the ferromagnetic quantum critical point. Previous application of the quantum order-by-disorder approach to the three-dimensional 
case\cite{andrewgareth,una} showed that quantum fluctuations not only drive the transition first-order at low temperatures, but also stabilize an incommensurate 
spiral phase below the tricritical point. These results are valid in the vicinity of the tricritical point but fail for much smaller temperatures, where the phase behavior is no longer 
controlled by the smallness of the magnetization $M$ but instead by the leading divergences of all orders in $M$ as $T\to 0$.

In this work we have presented 
a detailed analysis of the non-analytic structure of the fluctuation corrections to the free energy in $d=2$ and $d=3$. We have demonstrated that there exists an 
underlying hierarchy of divergences and obtained closed-form expression for the re-summation of the leading terms of all orders in $M$. By adopting the 
fermionic quantum order-by-disorder approach and self-consistently expanding around an electronic state with a spiral magnetization, we have derived re-summed expressions for the free energy of the spiral ferromagnet. This re-summation of leading divergences allows us to track phase boundaries at 
low temperatures far away from the tricritical point.

Our results demonstrate that not only the fluctuation-driven first-order transitions but also the instabilities towards spiral order are generic
features of itinerant ferromagnets. Despite the different non-analytic structures, we find very similar phase diagrams in $d=2$ and $d=3$. There are however differences
which can be tested by future experiments. In $d=3$, the spiral phase is found in a narrow region on the border of ferromagnetism, leading to a sequence of
transitions from paramagnet to spiral and finally to ferromagnet as temperature is decreased, consistent with recent experiments on PrPtAl (Ref.~\onlinecite{Jabbar+13}).
This order of transitions is not possible in $d=2$. Because of the re-entrant behavior of the phase boundary of the homogeneous ferromagnet, we predict that the spiral 
phase is located below the ferromagnetic state and stable over a larger region of the phase diagram. 

Expressions for non-analytic contributions to the free energy have been previously obtained by diagrammatic methods.\cite{belitz1,andre2, joe1, catherine1,joe2, andre1,belitzkirkpatrick} Our work shows that the diagrammatic calculations are equivalent to self-consistent second-order perturbation
theory. In  $d=3$,  we indeed recover the previously-known\cite{Belitz+99} result $\Delta F_1 \sim M^4 \ln (M^2 +T^2)$, and find the exact form of the $T=0$ free energy. This analytic form allows us to find the precise location of the quantum critical point for the ferromagnet in $3d$. 

In $d=2$, we break new ground. In contrast to the three-dimensional case, a straightforward calculation of a Landau expansion for the free energy where each 
coefficient of $M^{2n}$ is a simple function of $T$ is not possible. This signposts the fact that the system is intrinsically non-analytic, even in the vicinity of the 
tricritical point. Indeed there is a lack of consensus on the form of the quantum corrections we should expect from diagrammatic work.\cite{joe2,belitzkirkpatrick}
In Ref.~\onlinecite{belitzkirkpatrick}, Belitz and Kirkpatrick argue that the fluctuation contributions are of the form $\Delta F\sim M^2 \sqrt{M^2 +T^2}$, which is 
in agreement with our result $\Delta F_1  \sim M^2 T$ in the regime $T\gg M$. However, the result of Ref.~\onlinecite{belitzkirkpatrick} is based on the assumption 
that the singular behavior of the fluctuation integral for $T \ll M$ is cut off in the same way. Our work shows that this assumption is incorrect and 
that the asymptotic behavior for $T\ll M$ is given by $\Delta F_1 \sim M^3 \ln M$.  

Our results lay the groundwork for studying the multi-critical behavior of itinerant ferromagnets. In $d=3$, the re-summed form for the phase diagram has already been 
used to study superconducting instabilities mediated by magnetic fluctuations.\cite{CGP} In this instance, the intertwined magnetic spiral state and the 
superconducting instability resulted in the formation of a pair-density-wave state at very low temperatures. In two-dimensional systems such exotic states might be
stabilized at higher temperatures since quantum fluctuations become more important and the resulting non-analyticities are of a different form. Future applications 
of the fermionic quantum order-by-disorder approach include the study of multiple-band effects, orbital fluctuations, and the competition between 
nesting instabilities and fluctuation-driven phase formation.

\begin{acknowledgments}
We thank Andrew Berridge, Gareth Conduit, Gregor Hannappel, Chris Hooley, Una Karahasanovic, and Ed Yelland for helpful discussions and suggestions. This work was 
funded by the EPSRC under grant codes EP/H049584/1, EP/I031014/1 \& EP/I004831/2.
\end{acknowledgments}

\appendix

\section{Particle-hole density of states}

In order to evaluate the fluctuation integral $\Delta F(M)$ (\ref{fluctint}), we must first calculate the particle-hole densities of states $\tilde{\rho}(q,\epsilon)$ and 
$\Delta \rho(q,\epsilon)$, which are defined in Eq.~(\ref{phDOS}). We will do this separately for the cases $d=3$ and $d=2$. 

\subsection{Three dimensions}

\noindent
In $d=3$, and for $M=0$, $\tilde{\rho}$ takes the form
\be
\begin{split}
{\tilde{\rho}} (q, \epsilon) &= \int_{-1}^{1} d \cos \theta  \int_0^{\infty} \frac{2\pi k^2 d k}{(2 \pi)^3}  n \left( \frac{k^2}{2} + \frac{q^2}{8} - \frac{k q \cos \theta }{2} \right) \\
& \,\,\,\,\,\, \times \delta( \epsilon - k q \cos \theta ).
\end{split}
\ee

The integrals over $\cos \theta$ and $k$ may be done exactly. We are not so interested in the particle-hole densities themselves, but rather in their derivatives 
with respect to $\mu$, which enter the integrals $J_{n,n}$. For the $n$-th derivative we obtain
\begin{equation}
\partial^{n}_\mu {\tilde{\rho}} (q, \epsilon) = {\tilde{\rho}}^{(n)} (q, \epsilon) = \frac{1}{(2\pi)^2 q} \partial^{(n-1)}_\mu n \left[\phi^-(q,\epsilon)\right],
\label{rho3}
\end{equation}
where we have defined 
\be
\phi^\pm (q,\epsilon) =\frac12 \left( \frac{\epsilon}{q} \pm \frac{q}{2}\right)^2. 
\label{phipm}
\ee
Similarly, we find that
\begin{equation}
\Delta \rho^{(n)} (q, \epsilon) = \frac{1}{(2\pi)^2 q}  \partial^{(n-1)}_\mu \left\{n \left[\phi^+(q,\epsilon)\right] n\left[\phi^-(q,\epsilon)\right] \right\}.
\label{deltarho3}
\end{equation}

\subsection{Two Dimensions}

\noindent
In $d=2$, we must calculate the integral
\be
\begin{split}
{\tilde{\rho}} (q, \epsilon) &= \int_0^{2\pi} \,d\theta \int_0^{\infty} \frac{k d k}{(2 \pi)^2} \, n \bigl( \frac{k^2}{2} + \frac{q^2}{8} - \frac{k q \cos{\theta}}{2} \bigr)\\
&\,\,\,\,\,\,\, \times \delta( \epsilon - k q \cos{\theta}).
\end{split}
\ee
We carry out the $\theta$ integral to get
\be
{\tilde{\rho}} (q, \epsilon) = \frac{2}{(2 \pi)^2}  \int_0^{\infty} k\, d k \, \frac{n(\frac{k^2}{2} + \frac{q^2}{8} - \frac{\epsilon}{2} )}{\sqrt{k^2 q^2-\epsilon^2}} \theta (k^2 q^2 - \epsilon^2 ).
\ee
By a suitable change of variables, we reduce this to
\be
{\tilde{\rho}} (q,\epsilon) = \frac{\sqrt{2}}{(2 \pi)^2 q} \int_0^{\infty} \, \frac{dx}{\sqrt{x}} n[ x+\phi^-(q,\epsilon)] ,
\label{rho2}
\ee
where the functions $\phi^\pm (q,\epsilon)$ are defined as before (\ref{phipm}). An identical calculation for $\Delta \rho$ yields
\be
\Delta \rho (q,\epsilon) = \frac{\sqrt{2}}{(2 \pi)^2 q} \int_0^{\infty} \, \frac{dx}{\sqrt{x}} n[x+\phi^+(q,\epsilon)] n[ x+\phi^-(q,\epsilon)].
\label{deltarho2}
\ee

Although it is possible to perform the integrals (\ref{rho2}) and (\ref{deltarho2}) and write the results in terms of  poly-logarithmic functions, the expressions in integral form will 
prove more useful to us.

\section{Calculation of $\Delta F_1(M)$ in $d=3$.}

It is convenient to define a new function $\bar{n}(x)$ to keep our calculations uncluttered; this takes the form
\be
\bar{n}(x) = \frac{1}{1+e^{x/T}}
\ee
and is a Fermi function with chemical potential $\mu = 0$. The most divergent part of the integral $J_{n,n}$ in $d=3$ comes from
\be
\begin{split}
J_{n,n} &= \frac{2}{(2 \pi)^6} \int \, \frac{ dq \, d\epsilon_1 \, d\epsilon_2}{(\epsilon_1 + \epsilon_2)} \delta^{(n-2)}  [\phi^-(q,\epsilon_1)  - \mu] \\
& \,\,\,\,\,\,\times \delta^{(n-2)} [\phi^-(q,\epsilon_2)   -\mu] \bar{n}( \epsilon_2 ).
\end{split}
\ee

We integrate by parts $(n-2)$ times each with respect to $\epsilon_1$ and $\epsilon_2$ to find
\be
\begin{split}
J_{n,n} &= \frac{2}{(2\pi)^6} \int \, \frac{q^{2n-2}}{k_F^{2n-2}} dq \, d\epsilon_1 \, d\epsilon_2 \biggl( \frac{ \partial^{n-2}}{\partial \epsilon_1^{n-2}} \frac{ \partial^{n-2}}{\partial \epsilon_2^{n-2}} \frac{1}{\epsilon_1 + \epsilon_2} \biggr) \\
& \,\,\,\,\,\, \times \delta \left[ \epsilon_1 -\frac{q}{2} (q-2k_F) \right] \delta \left[ \epsilon_2 -\frac{q}{2} (q-2k_F) \right]  \bar{n} ( \epsilon_2 ),
\end{split}
\ee
and use the $\delta$-functions to integrate over $\epsilon_1$ and $\epsilon_2$, 
\be
J_{n,n} = \frac{2 (2n-4)! }{(2\pi)^6} \int_{-2k_F}^{\infty} \, d \delta q \frac{(2k_F)^{2n-2}}{k_F^{2n-2} (2k_F \delta q)^{2n-3}} \bar{n}( k_F \delta q),
\label{JNN}
\ee
where $\delta q = q-2k_F$ is the deviation of the particle-hole pair's momentum from $2k_F$, which is expected to be small. We simplify this equation by folding the integral around $\delta q=0$ and use the expression $\bar{n}(k_F\delta q) = \frac{1}{2} [1-\text{tanh}(k_F\delta q/2T)]$. The second term provides a cut-off for the lower limit of the integral, which now yields 
\be
J_{n,n} = \frac{-4(2n-4)! \nu_F^3}{(2 \mu)^{2n-2} } \int_{\frac{T}{2\mu}}^2 \frac{du}{u^{2n-3}},
\label{JNN2}
\ee
where we have made a simple change of variables, $u=k_F\delta q$, and defined $\nu_F = {k_F}/{2 \pi^2}$. Substituting this expression into Eq.~(\ref{deltaF1}),
we find
\be
\Delta F_1 = -64 g^2 \nu_F^3 \mu^2 \int_{\frac{T}{2\mu}}^2 du \overset{\infty}{\underset{n=2}{\sum}} \frac{(2n-4)! (-1)^n (2gM)^{2n}}{(2n)! u^{2n-3} (2 \mu)^{2n} }.
\ee
Taking sequential derivatives with respect to $M$, we simplify our expression significantly to obtain
\be
\begin{split}
\partial_M^4 \Delta F_1 &= - \frac{64 g^6 \nu_F^3}{\mu^2}   \int_{\frac{T}{2 \mu}}^2 du \overset{\infty}{\underset{n=2}{\sum}} \frac{(-1)^n (2gM)^{2n-4}}{u^{2n-3} (2\mu)^{2n-4} } \\
&= -\frac{64g^6 \nu_F^3}{\mu^2}  \int_{\frac{T}{2 \mu}}^2 \, du \frac{u}{g^2 M^2 + u^2}\\
&=\frac{32g^6 \nu_F^3}{\mu^2} \ln \left( \frac{g^2 M^2}{\mu^2} + \frac{T^2}{4 \mu^2} \right).
\end{split}
\ee

To get $\Delta F_1$, we integrate this four times with respect to $M$, keeping only terms that will give us an overall coefficient of $M^4$. This gives
the final closed form expression 
\begin{equation}
\Delta F_1(M,T) = \frac{4g^6 \nu_F^3 M^4}{3\mu^2} \ln \left(  \frac{g^2 M^2}{\mu^2} + \frac{T^2}{4 \mu^2} \right)
\end{equation}
for the re-summation of leading divergences in $d=3$.

\section{Calculation of $\Delta F(M)$ in $d=3$.}

We wish to use the recursion relation given in the text ~(\ref{recursion}) to calculate the complete, all-orders re-summation at $T=0$. Firstly, we write $\Delta F_1$ in terms of the variable $y = \sqrt{2} \mu / gM$, and define the functions $f_n(y)$ by

\be
\Delta F_n (M) = (gM)^2 f_n(y).
\ee

In terms of this new variable, the recursion relation ~(\ref{recursion}) takes the form

\be
f_n(y) = \frac{1}{n!} \partial_y^{2n-2} f_1 (y).
\ee

By Fourier transforming this expression to $k$-space, and after summation over $n$, we find that $\Delta F = \underset{n}{\sum} \Delta F_n$ is given by

\be
\Delta \tilde{F} (k) = g^2 M^2 \frac{ 1-e^{-k^2}}{k^2} {\tilde{f}}_1 (k).
\ee

We approximate the Fourier transform of $(1-e^{-k^2})/k^2$ by a triangular function with the same total area, to do the convolution and approximate for $y \gg 1$. Expressing the result in terms of $M$, we find the full resummation of the $T=0$ quantum corrections to all orders, which is given by

\be 
\Delta F (M) =  2 \sqrt{\pi} \lambda M^4 \ln \biggl[ \frac{ g^2 M^2}{\mu^2} \biggr].
\ee

\section{Calculation of $\Delta F_1(M)$ in $d=2$.}

Using the integral expressions (\ref{rho2}) and (\ref{deltarho2}) for the particle-hole densities of states in $d=2$, $J_{n,n}$ is given by a five-dimensional integral
\be
\begin{split}
J_{n,n} &= \frac{2}{(2\pi)^5}\int \, \frac{dq}{q}   d{\epsilon}_1 \, d{\epsilon}_2 \,dx \,dy \partial_\mu^n n [x + \phi^- (q,\epsilon_1) ]\\
& \,\,\,\,\, \times \frac{\partial_\mu^n \left\{ n [y + \phi^- (q,\epsilon_2) ] n [y + \phi^+(q,\epsilon_2)] \right\} }{\sqrt{x y}(\epsilon_1 + \epsilon_2)},
\end{split}
\ee
where $\phi^\pm (q,\epsilon)$ are defined as before and the ranges of integration are $\epsilon_1, \epsilon_1 \in (-\infty, \infty)$ and $q,x,y \in [0,\infty)$.

In order to do the integrals over $\epsilon_1$ and $\epsilon_2$, we first focus on the most divergent term, where all the derivatives hit the second Fermi function, and none 
hit the third. We make the same approximation as before, namely that $n^\prime (x) = -\delta (x-\mu)$, and then linearize the arguments of the derivatives of the Fermi functions 
as for $d=3$. We write the derivatives with respect to argument in terms of derivatives with respect to $\epsilon$, and then integrate by parts with respect to 
$\epsilon_1$ and $\epsilon_2$, $(n-1)$ times each. This gives
\be 
\begin{split}
J_{n,n}  &= \frac{2(2n-2)!}{(2 \pi)^5}\int \frac{dq}{q} \,d{\epsilon}_1 \, d{\epsilon}_2 \, dx \,dy \frac{  \bar{n}(\epsilon_2)}{\sqrt{x y}(\epsilon_1 + \epsilon_2)^{2n-1}}   \\
& \times \frac{q^{2n}}{k_F^{2n}}  \delta ( y -  k_F \epsilon_2/q + k_F \delta q ) \delta (  x - k_F \epsilon_1/q + k_F \delta q ).
\end{split}
\ee
After these approximations, the integrals over $\epsilon_1$ and $\epsilon_2$ are trivial and we obtain 
\be 
\begin{split}
J_{n,n}  &= \frac{2(2n-2)!}{(2\pi)^5 k_F} \int_{-2k_F}^\infty \, d\delta q \int_0^\infty dx \int_0^\infty dy   \frac{1}{\sqrt{x y}}\\
& \,\,\,\,\, \times \frac{\bar{n} (2y+ k_F \delta q)}{(x+y+2k_F \delta q)^{2n-1}}
\end{split}
\label{jnn}
\ee
with $\delta q = q-2k_F$ as before. We now feed this back into the re-summation expression (\ref{deltaF1}), and simplify by taking two $M$-derivatives, 
\be 
\begin{split}
\partial_M^2 \Delta F_1  &= \frac{-32g^4}{(2 \pi)^5 k_F} \int_{-2k_F}^\infty d\delta q \int_0^\infty dx \int_0^\infty dy \frac{1}{\sqrt{x y}}  \\
& \,\,\,\,\,\,\, \times \frac{ x+y + k_F \delta q}{ 4g^2 M^2 +(x+y+k_F \delta q)^2} \bar{n} (2y + k_F \tilde{q}). 
\end{split}
\label{d2ferromagnet}
\ee
Changing variables $u=y+k_F \delta q$ and doing the integral over $x$ we get
\be 
\begin{split}
\partial_M^2 \Delta F_1  &= \frac{-32g^4 }{(2 \pi)^5  k_F} \frac{\pi}{\sqrt{2}} \int_{-2k_F^2}^\infty du \int_0^\infty dy   \frac{1}{\sqrt{y}}\\
&\,\,\,\,\, \times \sqrt{ \frac{ \sqrt{u^2 +4g^2 M^2}+u}{u^2 + 4g^2 M^2}}  \bar{n} (y +u).
\end{split}
\ee
We now do the $y$-integral to get
\be 
\begin{split}
\partial_M^2 \Delta F_1 &= \frac{32g^4 \pi^{3/2} \sqrt{T} }{(2 \pi)^5 \sqrt{2} k_F} \int_{-4 \mu}^\infty du \, \text{Li}_{1/2}(-e^{-u/T}) \\
& \,\,\,\,\,\, \times  \sqrt{ \frac{ \sqrt{u^2 +4g^2 M^2}+u}{u^2 + 4g^2 M^2}}.
\label{df2d2}
\end{split}
\ee

Since this integration cannot be done analytically, we approximate it in two limits. Firstly, when $T \gg M$ we take $M=0$ inside Eq.~(\ref{df2d2}), which becomes
\be 
\partial_M^2 \Delta F_1  = \frac{32g^4 \pi^{3/2} \sqrt{2T} }{(2 \pi)^5 \sqrt{2} k_F}  \int_{-4 \mu}^\infty \frac{du}{\sqrt{u}}  \text{Li}_{1/2}(-e^{-u/T}) .
\ee
Only positive values of $u$ contribute to this integral when taking the limit $T \rightarrow 0$; the lower limit becomes $u=0$. We may then scale out $T$ to get
\be
\partial_M^2 \Delta F_1 = \frac{32g^4 \pi^{3/2} T }{(2 \pi)^5 k_F}  \int_0^\infty    \frac{du}{\sqrt{u}} \times \text{Li}_{1/2}(-e^{-u}) .
\ee
We split the integral into two regions, $u \in [0,1]$ where we approximate $\text{Li}_{1/2} (-e^{-u/T}) \approx \text{Li}_{1/2}(-1)$ and $u \in [1,\infty)$ where $\text{Li}_{1/2} (-e^{-u/T}) \approx 0$ to give 
\be
\partial_M^2 \Delta F_1 = \frac{64g^4 \pi^{3/2} T }{(2 \pi)^5 k_F} \text{Li}_{1/2} (-1),
\ee
which corresponds to a term in the free energy
\be 
\Delta F_1 (M) = \frac{16 \sqrt{2} g^4 \pi^{3/2}  \text{Li}_{1/2} (-1)}{(2 \pi)^5}  M^2 T.
\ee

Next, we turn our attention to the limit where $T \ll M $. Setting $T=0$ in Eq.~(\ref{df2d2}), we now see that only values of $u<0$ contribute. For this range of $u$, we may approximate 
the poly-logarithm by $\text{Li}_{n} (-e^u) \sim u^n / \Gamma (n+1)$ where $\Gamma$ is the Euler gamma function. This gives
\be 
\begin{split}
\partial_M^2 \Delta F_1 &= \frac{-32g^4 \pi^{3/2} }{(2 \pi)^5 \sqrt{2} k_F \Gamma (3/2)} \\
& \,\,\,\,\, \times \int_0^{4 \mu} du  \sqrt{ \frac{ u \sqrt{u^2 +4g^2 M^2}- u^2}{u^2 + 4g^2 M^2}}.
\end{split}
\ee
We rescale $u= 2 M \tilde{u}$, and scale out $\mu$ to get
\be 
\begin{split}
\partial_M^2 \Delta F_1  &=  \frac{-32g^4 \pi^{3/2} M }{(2 \pi)^5 \Gamma (3/2)} \\
& \times \int_0^{2/M} d\tilde{u}  \sqrt{ \frac{ \tilde{u} \sqrt{{\tilde{u}}^2 +g^2}- {\tilde{u}}^2}{{\tilde{u}}^2 + g^2}}.\\
\end{split}
\ee
Splitting the integral as before, and approximating the two halves we find 
\be
\begin{split}
& \int_0^{2/M} d\tilde{u}  \sqrt{ \frac{ \tilde{u} \sqrt{{\tilde{u}}^2 +g^2}- {\tilde{u}}^2}{{\tilde{u}}^2 + g^2}}\\
& \approx \int_0^{1} d\tilde{u}  \sqrt{ \frac{ \tilde{u} \sqrt{{\tilde{u}}^2 +g^2}- {\tilde{u}}^2}{{\tilde{u}}^2 + g^2}} + \int_1^{2/M} d\tilde{u} \frac{1} {\sqrt{{\tilde{u}}^2+g^2}} .
\end{split}
\end{equation}

The first term gives us a numerical constant, which corresponds to a sub-leading $\partial_M^2 \Delta F_1 \sim M$ piece. The second integral gives us
 $\text{ArcSinh}(2/M)-\text{ArcSinh}(1)$. Recall that $\text{ArcSinh}(x) = \ln (x + \sqrt{x^2+1})$, and that we are working in the limit where $M<1$, then this first, positive 
 term gives us a leading contribution 
\be 
\partial_M^2 \Delta F_1  =  \frac{32g^4 \pi^{3/2}}{(2 \pi)^5 \Gamma (3/2)} M \ln M,
\ee
which corresponds to the term 
\be 
\Delta F_1 (M) = \frac{16g^4 \pi^{3/2}}{(2 \pi)^5 \Gamma (3/2)} M^3 \ln M ,
\ee
in the free energy for sufficiently small values of $T$.\\


\begin{thebibliography}{43}
\bibitem[{doubfac()}]{doubfac}
\bibinfo{note}{The double factorial function is defined for odd argument as $(2k-1)!! = (2k)!/2^k k!$ and for even argument as $(2k)!!= 2^k k!$. }
\bibitem[{lif()}]{lifshitz}
\bibinfo{note}{Recall that we define the ``optimal'' $\tilde{Q}$ for a given
  $M$ as $\tilde{Q}(M)$ which solves the equation $\partial_{Q^2}
  \mathcal{F}(M,\tilde{Q}(M)) = 0$. Using Eq.~(\ref{qspiral}), we see that this
  translates into $\langle \eta_{\mathbf{k}}^2 F^\prime (M^2 +
  \eta_{\mathbf{k}}^2 \tilde{Q}^2) - \eta_{\mathbf{k}}^2 F^\prime (
  \eta_{\mathbf{k}}^2 \tilde{Q}^2) \rangle = 0$. Taking the limit that
  $\tilde{Q}(M) \rightarrow 0$ in this expression, and assuming that the
  angular average is non-zero, we find that this reduces to the expression
  $\partial_{M^2}F(M)|_{M=0} = 0$, which is the condition that $\alpha = 0$.
  Therefore, the Lifshitz line for the spiral coincides with the extension of
  the second-order line below $T_c$.}
\bibitem[{alph()}]{alpha}
\bibinfo{note}{By doing the re-summation for $T=0$, we remove one of the two energy scales in the problem completely. As a result, we cannot garner any information about the precise nature of the relative scaling between $M$ and $T$ that would be imposed on us by doing the complete re-summation for $T \neq 0$.}
\end{thebibliography}
\end{document}